**Title: The ecological and evolutionary energetics of hunter-gatherer residential mobility**

Running Title: Hunter-gatherer residential mobility

Author 1: Marcus J. Hamilton[1,2] (corresponding author)

Author 2: José Lobo[3]

Author 3: Eric Rupley[1,4,5]

Author 4: Hyejin Youn[1,6,7]

Author 5: Geoffrey B. West[1]

[1] Santa Fe Institute, 1399 Hyde Park Rd, Santa Fe, NM 87501 USA

[2] School of Human Evolution and Social Change, Arizona State University, Tempe, AZ 85287 USA

[3] School of Sustainability, Arizona State University, Tempe, AZ 85287 USA

[4] Department of Anthropology, University of Michigan, Ann Arbor, MI 48109 USA

[5] Museum of Anthropological Archaeology, University of Michigan, Ann Arbor, MI 48109 USA

[6] Institute for New Economic Thinking at the Oxford Martin School, University of Oxford, OX2 6GG, UK

[7] Mathematical Institute, University of Oxford, Oxford, OX2 6GG, UK

**Key Words**

Spatial ecology; metabolic theory of ecology; body size; temperature-dependence; net primary production

**About the authors**

Marcus Hamilton is an evolutionary anthropologist and archaeologist who uses formal ecological theory as a baseline to study human ecological predictability and uniqueness. He is a Postdoctoral Fellow at the Santa Fe Institute and Arizona State University. José Lobo is a Research Faculty at Arizona State University's the School of Sustainability. An urban economist, his research spans the origins of human organizations are different scales. Eric Rupley is an archaeologist, who studies the long-term evolution of human systems, currently a Postdoctoral Fellow at the Santa Fe Institute. Hyejin Youn is a statistical physicist interested in the dynamics of social systems, currently a Senior Research Fellow at Oxford University. Geoffrey West is a Distinguished Professor at the Santa Fe Institute. He is a theoretical physicist, specializing in building formal theory in biological and social systems.

## Introduction

Residential mobility is a key aspect of hunter-gatherer foraging economies, and is therefore an issue of central importance in hunter-gatherer studies. [1-7] Hunter-gatherers vary widely in annual rates of residential mobility, and understanding the sources of this variation has long been of interest to anthropologists and archaeologists. The vast majority of hunter-gatherers dependent on terrestrial plants and animals move camp multiple times a year because local foraging patches become depleted, and food, material, and social resources are heterogeneously distributed through time and space. In some environments, particularly along coasts, where resources are abundant and predictable, hunter-gatherers often become effectively sedentary. But even in these special cases, a central question is how these societies maintained a viable foraging economy while reducing residential mobility to near zero.

While the causes of hunter-gatherer mobility undoubtedly include a combination of cultural, economic, and biological factors, here, we focus on the coarse-grained ecological and energetic constraints of residential mobility. We define hunter-gatherer residential mobility as the pattern of camp-to-camp movements of individuals within a population over the course of a year in order to effectively exploit their environments so as to meet their nutritional, material, and social requirements. Residential mobility is commonly measured as the total distance moved per year, and the number of moves made per year. An important metric of interest is thus the average distance per residential move, particularly as mobility decisions are commonly made on a short-term basis, rather than annually.

**While residential mobility is fundamental to the hunter-gatherer lifeway, it is extremely costly in terms of energy** (the basic metabolic cost of movement)**; logistics** (the successful planning and organization of moving individuals, families, and their material culture)**; opportunity loss** (time spent moving camps is time not spent foraging, child-rearing, producing tools, or any of the many other tasks contributing to fitness)**, and time itself** (there is a strictly finite number of days in the year). Given these costs, residential mobility constitutes a large fraction of a hunter-gatherer energy budget (or opportunity/logistic/time budget). It is therefore reasonable to assume that these costs should be minimized within the constraints of local ecosystems.

The central question we pursue here is: how predictable is variation in hunter-gatherer residential mobility? We address this using ethnographic data, with which we assess a set of hypotheses, derived from formal theory[8-10], that make specific predictions about the scale and variation of hunter-gatherer residential mobility. These predictions are based on 1) the evolved biomechanics and bioenergetics of humans; and 2) the turnover of energy in ecosystems. In addition, an implicit question we also address is whether hunter-gatherer mobility is predictable from what is known of evolutionary and ecological constraints on mammalian mobility.

Our framework jointly considers two activities that are central to hunter-gatherer spatial ecology: 1) use of space; where space is taken to be the resource catchment area for food, water and all material and social resources needed for survival; and 2) movement in space; the set of constraints that shape the effective utilization of space and the energetic limits of body size. Our goals are to link the use of, and movement in, space in a formal conceptual framework, and show empirical support for the predictions generated by such a framework using independent data sets on hunter-gatherer area use and mobility. Our model explains and predicts the variation in observed residential mobility. These results can be extended both to the past, and the present, for cases where data are not currently available. Specifically, we show that both the scale of hunter-gatherer mobility, and variation in rates of mobility across different environments can be mathematically derived from fundamental ecological theory. We present a conceptual flow chart of our theoretical approach in Figure 1.

**Theoretical framework**

Since mobility is, to a large extent, driven by the need for a continuous supply of energy in the form of food resources, a natural framework for addressing this question is provided by the metabolic theory of ecology[8,10]. This body of theory provides a powerful framework for formulating formal testable hypotheses concerning evolutionary and ecological constraints on the scale and variation of hunter-gatherer residential mobility. It is based on the observation that energy supply and utilization are fundamental to all biological and ecological processes, that "metabolism is to ecology as genetics is to evolution"[8]. Understanding how energy fluxes through organisms (i.e., metabolism), and how this scales across units of biological organization (which are essentially collections of organisms, such as populations, communities, ecosystems, and the biosphere) is as fundamental to understanding ecology as understanding how genetic processes operating at the same scales of biological organization is to evolution.

The metabolic theory of ecology is based on the empirical observation (known as Kleiber's Law) that the basal metabolic rate, $B$, of an organism scales with its mass, $M$, as a simple power law:

$$B = B_0 M^{\alpha} \qquad (1)$$

where $B_0$ is a taxon-specific constant, and $\alpha \approx 3/4$ . Similar scaling laws hold for almost all physiological traits and life-history events across all taxonomic groups, manifesting exponents that are typically simple multiples of ¼. The origin of these ubiquitous 1/4-power scaling laws arises from the fractal-like, space-filling nature of optimized internal networks that distribute the energy and resources that sustain organisms (such as the vascular system).[11-13] The overall scale parameter, $B_0$, is derived from the underlying biochemical kinetics of metabolism and, as such, depends on the temperature, $T$, at which an organism operates (14). This is given by the exponential Arrhenius-Boltzmann factor, $\exp(-E/kT)$, where $E$ (~0.6 eV K-1) is the average activation energy of the biochemical reactions contributing to metabolic processes, $k$ is Boltzmann's constant $(8.62 \times 10^{-5}$ eV K-1$)$ , and $T$ is the absolute temperature at which the organism operates (ºK). Incorporating this into equation 1 yields[14]

$$B = B_0 e^{-E/kT} M^{\alpha} \qquad (2)$$

where $B_0$ is a mass and temperature independent constant. For endotherms, $T$ is the internal body temperature, and for ectotherms, $T$ is the ambient environmental temperature. A large body of empirical work supports equation 2 as a fundamental description of the metabolic rate of individual plants and animals. Essentially, it states that the metabolic rate of an organism is a function of its size, the structure of its internal distribution networks, and the temperature at which it operates. Put slightly differently, it expresses the fact that mass and temperature are the major determinants of the variation of physiological traits and life history events and that these are encapsulated in just two numbers: ¼, derived from network dynamics, and E ~ 0.6 eV, the scale of fundamental biochemical reactions.

As ecosystems are composed of individual organisms it follows that the respiration (or metabolism) of an entire ecosystem, $R$, is the sum over all individuals, $B_i$ :[15]

$$R = \sum_i B_i \qquad (3)$$

where $i$ indexes individuals. Because the vast majority of biomass in an ecosystem is composed of microbes and plants, both of which are ectothermic, it follows that the temperature-dependence of the metabolic rate of an ecosystem is primarily governed by a single Arrhenius-Boltzmann factor:

$$R \propto e^{-E/kT} \qquad (4)$$

where $T$ is the ambient environmental temperature. Importantly, because equation 4 describes the temperature-dependent metabolic flux of an ecosystem, all biotic interaction rates associated with metabolism, such as biomass growth, disease load, and predator-prey interaction rates are predicted to exhibit the same temperature-dependence.[8]

In addition to the environmental temperature, there are other important rate-limiting constraints on $R$, such as water-availability.[8,9] For example, not all warm ecosystems have high rates of biomass turnover (e.g., warm deserts), whereas some do (e.g., the tropics). Therefore, we introduce annual precipitation, $P$ (mm/yr), as the rate-limiting constraint into equation 4, giving the expression

$$R \propto P^{\theta} e^{-E/kT} \qquad (5)$$

where the exponent $\theta$ captures the response of a change in resource supply rate to a change in precipitation.

In the rest of the paper, we use equation 1 to predict the overall scale of hunter-gatherer residential mobility based on human body size, and equation 5 to predict the variation in rates of residential mobility across hunter-gatherer cultures based on ecosystem energetics.

**Data**

We use three data sets. The first is terrestrial mammal migration data from Hein et al.[16] The data is in terms of two variables ($n$ = 33):

1. Mammalian body size ($M$): the average body size of a species (in kg).
2. Annual migration distance ($D$): the total migration distance covered by an individual of that species per year (in km).

The sample size is relatively small due to the paucity of high quality data.

The second data set is a collection of estimates (n = 47) of the total distance travelled per year ($D$) by hunter-gatherer groups, from Kelly.[1]

The third data set is hunter-gatherer spatial ecology using data from Binford.[3] The data we use consist of seven variables and includes all groups that move at least once per year ($n$ = 314):

1. Area ($A$): the total territory size used by a population of hunter-gatherers over a year.
2. Population size ($N$): the total population size in an ethnolinguistic group.
3. Distance travelled per year ($D$): the total distance moved over the course of the year due to residential movements from patch to patch (in km).
4. Number of moves ($V$): the total number of residential moves made in a year.
5. Average annual temperature ($T$): the average environmental temperature of the population.
6. Precipitation ($P$): the average annual rainfall experienced by the population (in mm/yr).
7. Net primary production ($NPP$): the net production of biomass produced over the course of a year in an ecosystem (in $g/m^2/yr$ i.e., the annual turnover of biomass). $NPP$ is therefore a measure of the flux of free energy in an ecosystem.

The Binford data is the largest, and most recent, cross-cultural collection of data on global variation in hunter-gatherer ecology currently available. The data were collated primarily from primary sources in the ethnographic and ethnohistoric literature, and so, like any dataset, is subject to errors introduced during collection and collation. However, assuming data assembly was not systematically and simultaneously biased across multiple variables, the sources of error should be statistically independent. The predominant effect should therefore be to introduce noise into the analysis but without significantly affecting underlying statistical trends. The Kelly data set is similar to the Binford data, collected in similar ways, and so is likely subject to the same sources of error.

**Evolutionary constraints on the scale of hunter-gatherer mobility**

Mobility is a fundamental component of the ecology of most animals, as food resources are heterogeneously distributed in time and space, and energy storage capacity is strictly limited.[17,18] This, of course, is the case for most mobile hunter-gatherers where local resource patches become depleted over time, and the capacity to store food in most environments is limited to no more than a few days without effective bulk food preservation techniques. While traditional storage technologies such as drying or smoking meat are (and were) widely used, their effectiveness is limited, and rarely able to support the total energy requirements of local populations for any substantial amount of time, except for in a very limited set of circumstances. For example, hunter-gatherers can become effectively sedentary (or semi-sedentary) in locally-specific environments, especially along coast or lake shores, where resources are abundant, predictable, and storable. However, this is rarely the case for terrestrial foragers, the vast majority of which are nomadic and residentially mobile.

**Because mobility is energetically costly it is always minimized given the biomechanical and bioenergetic constraints of an organism**.[16-18] This minimization principle is well-known in mammals whose primary mode of migration is walking, where total annual migration distance, $D$, scales with body size $M$, as[16]

$$D = d_0 M^{1/3} \tag{6}$$

where $d_0$ is a constant capturing the ratio of the capacity of mammals to store energy to the metabolic cost of transport, both of which scale with body size. The 1/3-scaling with body size results from the geometric principle that stride length is a linear dimension, and so scales as a 1/3-power of a volume (i.e., body size).

Given that: 1) humans are mammals; 2) mobile hunter-gatherers have extremely limited abilities to store significant amounts of energy beyond body fat; and 3) local resources become depleted requiring residential movements to new patches[19], we hypothesize that equation 6 describes the average total distance moved per year across our sample of hunter-gatherer societies. Specifically, we test the hypothesis that the total distance moved per year by the average terrestrial hunter-gatherer group is consistent with that predicted by our body size. We examine this using mammal body size and migration distance, as described above, against which we plot the average ($\pm$ 2 s.d.'s) total distance moved per year as a function of average hunter-gatherer body size.

Figure 2 shows that total distance moved per year by hunter-gatherers (using both the Kelly and the Binford data sets) is remarkably close to the expected annual migration distance for a 60 kg mammal. This demonstrates that the overall scale of mobility exhibited across hunter-gatherer societies is a function of the evolutionary biomechanics and bioenergetics of human body size. While human body size sets the overall scale of hunter-gatherer annual mobility, we now turn to understanding variation in rates of residential mobility observed across hunter-gatherer populations.

**Ecological constraints on the variation of hunter-gatherer mobility**

Over human evolutionary history, hunter-gatherers have existed in the vast majority of terrestrial environments across the planet[1]. Hamilton et al.[20] showed that size of a hunter-gatherer territory varies predictably with coarse-grained environmental constraints, specifically, environmental temperature and water availability. This is an important result because the total size of a territory is the sum of individual home ranges, $H$, which are defined as the area used by an individual of body size, $M$, to meet its metabolic requirements, $B$, given the resource supply rate (or energy availability) per unit area, $R$, of the local environment. Effectively, a home range is a resource catchment area. Home range is thus defined as $H_0 \equiv B/R$ and a territory size, $A$, consisting of $N$ individuals is

$$A = \sum_i^N H_{0,i} \qquad (7)$$

Because humans do not differ in body sizes by orders of magnitude, we hold $M$, and consequently $B$, fixed, corresponding to an average mass of 60kg (see[21] for further discussion of variation in hunter-gatherer body sizes). Hamilton et al.[20] also showed that hunter-gatherers exhibit *economies of scale* in their spatial ecology, reflected in the area per individual, $A/N$, decreasing predictably with population size. Here, the term "economies of scale" refers to the cost or efficiency advantages accrued with increased output or level of activity. Because $A = H_0 N^\beta$, where $\beta \approx 3/4$, the area per individual scales as[20]

$$A/N = H_0 N^{-1/4} \qquad (8)$$

Equation 8 captures an economy of scale because it quantifies the energetic benefits to an individual from existing within a population of $N$ individuals. Assuming that home ranges are constant, body size is not density dependent[18] and that there is no significant improvement in extractive technologies, equation 8 implies that home ranges effectively overlap with larger populations $N$. This results in an effective reduction in the exclusive area an individual requires to meet their metabolic requirement, $A/N$, at a rate proportional to $N^{-1/4}$.[20,22]

As stated above, variation in hunter-gatherer space-use is constrained by the average environmental temperature and water availability within territories. Assuming nutrient availability [23], temperature and water availability yields a Net Primary Production (*NPP*) which can be defined as the density of biomass produced per year (g/m²/yr). Because *NPP* is measured in units of time and area, it is the annual flux of useful chemical energy (Gibbs free energy) in an ecosystem that is used to produce new biomass. Indeed, a multiple regression of $\ln NPP = f\left(1/kT, \ln P\right)$ from hunter-gatherer territories demonstrates that 83% of the variation in *NPP* is explained by these two variables (see Table 1 for regression results). Figure 3 is a contour plot showing *NPP* is highest in warm/wet hunter-gatherer territories (i.e., the tropics) and lowest in cold/dry environments (i.e., arctic/tundra). This is important for hunter-gatherer ecology because biodiversity increases monotonically with *NPP*[24,25], and therefore so does the potential diversity of harvestable plants and animals.[1]

However, in the following we are interested in measuring the individual effects of temperature and water availability on hunter-gatherer spatial ecology, and so treat the two components of *NPP* individually. As such, the environmental constraints on hunter-gatherer home range are described by:

$$H_0 = \frac{\langle B \rangle}{R} = \frac{\langle B \rangle}{c_0 P^\theta e^{-E/kT}} \tag{9}$$

Where $c_0$ is a constant. We then write a full expression of the scaling of hunter-gatherer territory sizes as

$$A = \langle B \rangle c_0^{-1} P^{-\theta} e^{E/kT} N^\beta \tag{10}$$

The area per individual is therefore given by

$$A/N \propto P^{-\theta} e^{E/kT} N^{\beta-1} \tag{11}$$

The above expression describes the environmental- and population size-dependence of the area required by an individual to meet their annual metabolic requirements. Having established the environmental constraints on individual space-use by hunter-gatherers[20,22], we now turn to modeling the mobility strategy employed to access that space.

By definition, mobile hunter-gatherers divide the total area they require to meet their annual energy budget into a discrete number of annual moves between patches. Under the assumption that hunter-gatherers will minimize the energetic costs of accessing this area we hypothesize that, on average, rates of hunter-gatherer mobility should show the same temperature dependence as the area use described above. That is to say, two fundamentally inter-dependent aspects of hunter-gatherer spatial ecology are home range size and mobility, and as such, both should show the same environmental-dependence.

A mobile hunter-gatherer population moves $V$ times per year covering a total distance $D$ over a year. Therefore, the average distance per move (i.e., distance between residential patches) is $\langle D \rangle \equiv D/V$. Given equation 11 we hypothesize that $\langle D \rangle = f(T, P, N)$. Figure 4 shows that

$$\langle D \rangle \propto \left( \frac{A}{N} \right)^{1/4} \tag{12}$$

Combining equations 11 and 12 leads to the prediction that

$$\langle D \rangle \propto P^{-\frac{1}{4}} e^{\frac{E}{4}\frac{1}{kT}} N^{-\frac{1}{16}} \qquad (13)$$

Because the scaling of $\langle D \rangle$ with $N$ is so shallow (with a scaling exponent of only -1/16), for small hunter-gatherer populations its value is effectively zero. We test this prediction using the hunter-gatherer annual mobility data from Binford[3] described above. Consistent with our predictions, Figure 5 shows that average distance per move is both temperature and precipitation dependent. More accurate estimates of the scaling exponents result from a regression model that captures the simultaneous effects of temperature, rainfall and population size on average travel distance (see Table 2 for results). Again, consistent with our predictions (equation 14), results show that

$$\langle D \rangle \propto P^{-0.27} e^{0.15\frac{1}{kT}} N^{0.01} \qquad (14)$$

Therefore, not only is average distance per move temperature-dependent, but at the same rate as area-use, and this has been shown using independent data-sets.

**Discussion**

In this paper we showed that the average distance a hunter-gatherer band moves between patches is a direct function of the available energy in the local environment. Moreover, and importantly, we showed it is possible to quantitatively predict how hunter-gatherer mobility should vary across different environments given the fundamental biochemistry and kinetics of energy turnover in ecosystems. As predicted by ecological theory, the average distance between patches increases exponentially with decreasing temperature and with decreasing precipitation in response to decreasing flux of energy in ecosystems.

In specific, we considered two hypotheses concerning hunter-gatherer rates of residential mobility derived from the metabolic theory of ecology. The first states that the overall scale of hunter-gatherer annual residential mobility is determined by our species body size, due to biomechanical, bioenergetic, and geometric constraints. Ethnographic data suggest that, indeed, average levels of hunter-gatherer annual mobility are close to the predicted level of mobility for a 60kg mammal given the limited capacity to store energy. The second set of predictions states that since hunter-gatherer rates of mobility are responsive to energy availability in ecosystems, they are dependent on water-availability and the flux of energy in ecosystems. These rates of mobility are predicted by metabolic theory and are well supported by available data. We show that **the overall scale of hunter-gatherer mobility is set by human evolutionary biomechanics and bioenergetics, and that the observed variation in the rates of hunter-gatherer residential mobility across cultures is largely a function of energy availability in ecosystems.**

Given the well-known positive temperature-dependence of biodiversity (measured as the abundance of species per unit area, including plants, mammals, reptiles, amphibians, and birds[8,10,24-] [25]), our results suggest that hunter-gatherer mobility decreases as the potential diversity of prey species increases. Moreover, our results suggest that because biotic interaction rates are temperature-dependent, foraging interaction rates with prey species also increase with temperature and precipitation. That is to say, in warmer and wetter environments there will be a predictably higher diversity of potential prey species (see[26]), and therefore a higher interaction rate with harvestable resources.

However, there are also negative impacts of temperature on hunter-gatherer mobility. The same theory also predicts that warmer (and especially wetter) environments will have a greater disease interaction rate, as human pathogen load increases predictably with ecosystem temperature[27]. Combined with pathogen load and the fact that decay rates will also increase predictably with increasing temperature, limiting the ability to store food resources, disease avoidance and storage capabilities will serve to increase rates of residential mobility in warm and wet environments. Indeed, in our dataset, the eight groups that report the highest number of moves per year are all tropical foragers (and in some cases these estimates may be considerable underestimates, e.g. the Ache[28] and the Nukak[29]).

One way to visualize the remarkable empirical congruence between residential mobility and the availability of energy in ecosystems is to compare the contour plots of net primary production, Figure 3, and average distance per move, Figure 6, visually. These plots show that in two dimensions (temperature and precipitation) there is a remarkable similarity between *NPP* and averaged distance moved: Average distance moved is highest in regions of lowest *NPP* and decreases as *NPP* increases along very similar 2-dimensional gradients. These plots confirm that rates of hunter-gatherer mobility track gradients in the flux of energy in ecosystems. The bottom panel of Figure 6 shows that there is no clear structure to the residuals of average distance, indicating the relationship between mobility, temperature and precipitation is equally predictive across the 2-dimensional plot.

The results shown here are consistent with data and analysis reported in Kelly[1] where the average distance per move in hunter-gatherer groups decreases exponentially with effective temperature. However, because of the way temperature is reported in[1] as "effective temperature", these data cannot be used to validate our theory directly. Our results are also consistent with Binford's "packing model",[3] where hunter-gatherer mobility decreases with population density (holding environmental variation constant). For example, rearranging equation 12 we find $\langle D \rangle \propto (N/A)^{-1/4}$, which states that the average distance per move decreases with population density to the one quarter-power. This is particularly interesting as the naïve geometric expectation would be that a distance traveled should scale as the square root of the area covered (i.e., $L \propto A^{1/2}$). However, the ¼-power scaling relation between $\langle D \rangle$ and $A$ exhibited in the data (Figure 4 and equation 12) indicates an unexpected economy of scale in residential mobility, similar to that shown for area use (equation 8). Therefore, this observation suggests a further mechanistic connection between hunter-gatherer space-use and mobility, where the economy of scale in spatial energy use may result from the economy of scale in the mobility strategy employed to access that space. It is also interesting to note that from our results (equation 14), the average distance per move is effectively independent of population size (i.e., $\langle D \rangle \propto N^0$). This has two important implications: 1) the density-dependence of mobility is not a function of the number of people in the overall population, but in the amount of area they require to meet their requirements; and therefore 2) in those cases in prehistory where terrestrial hunter-gatherers became sedentary prior to the full development of agriculture, such as the Natufian Near East,[30] the primary cause of sedentism was likely changes in their subsistence ecology (effectiveness at utilizing the spatial availability of energy), not because of increasing population sizes *per se*.

While the hunter-gatherer dataset used in this paper is both large and geographically extensive, it is, by construction, exclusively historic in nature, limited to those hunter-gatherer societies for which written records exist, in either the present or recent past (a.k.a. the "ethnographic present"). Therefore, it is reasonable to be concerned that this is a biased sample of hunting and gathering as a human lifestyle *in toto* as we have fragmentary archaeological evidence for either prehistoric nor pre-agricultural hunter-gatherers. However, the strength of the explanatory approach in this paper, as opposed to ethnographic analogy[31], is that by understanding how underlying dynamics derived from general principles constrain observed distributions, we are able to make robust inferences about societies for which we have little or no data. Moreover, there is no *a priori* reason to believe that rates of residential mobility (and area use) in pre-historic or pre-agricultural hunter-gatherer societies would not have responded to fundamental ecological constraints, such as the availability of energy on landscapes, which we see in ethnohistoric data. Indeed, the contour map, Figure 6, provides predictions for the average distance an individual in a hunter-gatherer population could be expected to move, given any combination of environmental temperature, and annual precipitation, irrespective of time period. As such, the model provides novel quantitative insights into the conditions under which terrestrial hunter-gatherers may become sedentary. The predictive nature of this approach to understanding major transitions in human energetics will be the subject of future work.

Our results provide a statistically robust mechanistic explanation of how large-scale ecological constraints fundamentally shape the movement of hunter-gatherers through space as they utilize available energy. However, importantly, there is unexplained variation in our models. Sources of this unexplained variation likely include cultural, economic, and ecological processes, historical contingency, as well as measurement and statistical error. Therefore, our claim is not that all aspects of hunter-gatherer residential mobility (and spatial ecology as a whole) are simply responses to energy availability: the ways in which cultural mechanisms impact hunter-gatherer spatial ecology are of great interest, but are much harder to measure and not explicitly considered here. However, the overall constraints that shape hunter-gatherer spatial ecology are regular, predictable, and coarse-grained aspects of the environment.

## Acknowledgements

We thank Jim Brown, Briggs Buchanan, Paul Hooper, and Chris Kempes for commenting on previous versions of this manuscript. This research was partially funded by grants from the John Templeton Foundation Institute (MJH, JL, ER and GBW) and the National Science Foundation, grant SMA-1312294 (HY). GBW thanks the Eugene and Clare Thaw Charitable Trust for their generous support. This paper is a product of the "Universals in Human Biosocial Organization" working group at the Santa Fe Institute.

## References

1. Kelly R. 1995. The Foraging Spectrum. Washington. Smithsonian Institution Press.

2. Kelly R. 1983. Hunter-gatherer mobility strategies. J Anthropol Res 39:277-306.

3. Binford LR. 2001. Constructing Frames of Reference Berkeley, University of California Press.

4. Binford LR. 1981. Willow smoke and dog's tails: hunter-gatherer settlement systems and archaeological site formation. Am Antiq 45: 4-20.

5. Lee RB, DeVore I. 1986. Man the Hunter. New York, Aldine.

6. Grove M. 2009. Hunter-gatherer movement patterns: causes and constraints. Journal of Anthropological Archaeology 28: 222-233

7. Grove M. 2010. Logistical mobility reduces subsistence risk in hunting economies. Journal of Archaeological Science 37: 1913-1921

8. Brown JH, Gillooly JF, Allen AP, Savage, VM, West GB. 2004. Toward a metabolic theory of ecology. Ecology 85: 1771-1789.

9. Karasov WH, Martinez del Rio C. 2007. Physiological Ecology: How Animals Process Energy, Nutrients, and Toxins. Princeton, Princeton University Press.

10. Sibly RM, Brown JH, Kodric-Brown A eds. 2012. Metabolic Ecology: A Scaling Approach. : Oxford, Wiley-Blackwell.

11. West GB, Brown JH, Enquist B. 1997. A general model for the origin of allometric scaling laws in biology. Science 276: 122-126.

12. West GB, Brown JH, Enquist B. 1999. The fourth dimension of life: fractal geometry and allometric scaling of organisms. Science 284: 1677-1679.

13. West GB, Brown JH, Enquist B. 2001. A general model for the structure and allometry of plant vascular systems. Nature 413: 628-632.

14. Gillooly J, Brown JH, West GB, Savage VM, Charnov EL. 2001. Effects of size and temperature on metabolic rate. Science 293: 2248-2251.

15. Enquist, BJ, Economo EP, Huxman TE, Allen AP, Gillooly JF. 2003. Scaling metabolism from organisms to ecosystems. Nature 423: 639-642.

16. Hein A, Hou C, Gillooly JF. 2012. Energetic and biomechanical constraints on animal migration distance. Ecol Lett 15: 104-110.

17. Peters RH. 1983. The Ecological Implications of Body Size. Cambridge, Cambridge University Press.

18. Schmidt-Nielsen K. 1984. Scaling: why is animal size so important? Cambridge, Cambridge University Press.

19. Charnov EL. 1976. Optimal foraging, the marginal value theorem. Theor Popul Biol 92: 129-136.

20. Hamilton MJ, Milne BT, Walker RS, Brown JH. 2007. Nonlinear scaling of space use in human hunter-gatherers. Proc Natl Acad Sci USA 104: 4765-4769.

21. Walker R.S , Hamilton MJ. 2008. Life history consequences of density-dependence and the evolution of human body sizes. Curr Anthropol 491: 115-122.

22. Hamilton MJ, Burger O, DeLong JP, Walker RS, Moses ME, Brown JH. 2009. Population stability, cooperation and the invasibility of the human species. Proc Natl Acad Sci USA 10630: 12255-12260.

23. Anderson-Texeira K, Vitousek VM. 2012. Ecosystems. In Sibly RM, Brown JH, Kodric-Brown A, editors. Metabolic Ecology: a Scaling Approach. Oxford Wiley-Blackwell, p112-119.

24. Allen AP, Gillooly JF, Brown JH. 2002. Global biodiversity, biochemical kinetics, and the energetic equivalence rule. Science 297: 1545-1548.

25. Storch, D, Marquet PA, Brown JH, eds. 2007. Scaling Biodiversity. Cambridge, Cambridge University Press.

26. Hatton, IA, McCann, KS, Fryxell, JM, Davies, TJ, Smerlak, M, Sinclair, ARE, Loreau, M. 2015. The predator-prey power law: Biomass scaling across terrestrial and aquatic biomes. Science, 349: aac6284.

27. Fincher CL, Thornhill R. 2008. Assortative sociality, limited dispersal, infectious disease and the genesis of the global pattern of religion diversity. Proc R Soc Lond B: Biol Sci, 275: 2587-2594.

28. Hill KR, Hurtado AM. 1996. Ache Life History: The Ecology and Demography of a Foraging People. New York, Aldine de Gruyter.

29. Politis G. 2009. Nukak: Ethnoarchaeology of an Amazonian People. New York, Left Coast Press.

30. Simmons AH. 2011. The Neolithic Revolution in the Near East: Transforming the Human Landscape.  Tucson, University of Arizona Press.

31. Kelly R. 1999. Thinking about prehistory. In Beck C, editor. Models for the Millennium: Great Basin Anthropology Today. Salt Lake City, University of Utah Press, p111-117.

**Tables**

ANOVA table

| Source | DF | Adj SS | Adj MS | F-Value | P-Value |
|---|---|---|---|---|---|
| Regression | 2 | 349.47 | 174.74 | 818.84 | 0.000 |
| $E$ | 1 | 18.00 | 18.002 | 84.36 | 0.000 |
| $\ln P$ | 1 | 196.35 | 196.35 | 920.12 | 0.000 |
| Error | 336 | 71.70 | 0.21 | | |
| Total | 338 | 421.17 | | | |

| R-sq | R-sqadj) |
|---|---|
| 82.98% | 82.87% |

Coefficients

| Term | Coef | SE Coef | T-Value | P-Value |
|---|---|---|---|---|
| Constant | 6.841 | 0.84 | 8.13 | 0.000 |
| $E$ | -0.16 | 0.02 | -9.18 | 0.000 |
| $\ln P$ | 0.95 | 0.03 | 30.33 | 0.000 |

Table 1. Regression statistics and coefficients for Net Primary Production $\ln NPP$) as a function of temperature $1/kT$) and precipitation $\ln P$).

ANOVA table

| Source | DF | Adj SS | Adj MS | F-Value | P-Value |
|---|---|---|---|---|---|
| Regression | 3 | 44.98 | 14.99 | 73.89 | 0.000 |
| $E$ | 1 | 9.47 | 9.47 | 46.69 | 0.000 |
| $\ln P$ | 1 | 11.21 | 11.21 | 55.25 | 0.000 |
| $\ln N$ | 1 | 0.06 | 0.06 | 0.29 | 0.593 |
| Error | 234 | 47.48 | 0.20 | | |
| Total | 237 | 92.46 | | | |

| R-sq | R-sqadj) |
|---|---|
| 48.65% | 47.99% |

Coefficients

| Term | Coef | SE Coef | T-Value | P-Value |
|---|---|---|---|---|
| Constant | -1.72 | 0.983 | -1.76 | 0.080 |
| $E$ | 0.15 | 0.02 | 6.83 | 0.000 |
| $\ln P$ | -0.27 | 0.04 | -7.43 | 0.000 |
| $\ln N$ | 0.01 | 0.02 | 0.54 | 0.593 |

Table 2. Regression statistics and coefficients for average distance per move $\ln \langle D \rangle$ as a function of temperature ($E$), precipitation ($\ln P$), and population size ($\ln N$).

**Figures**

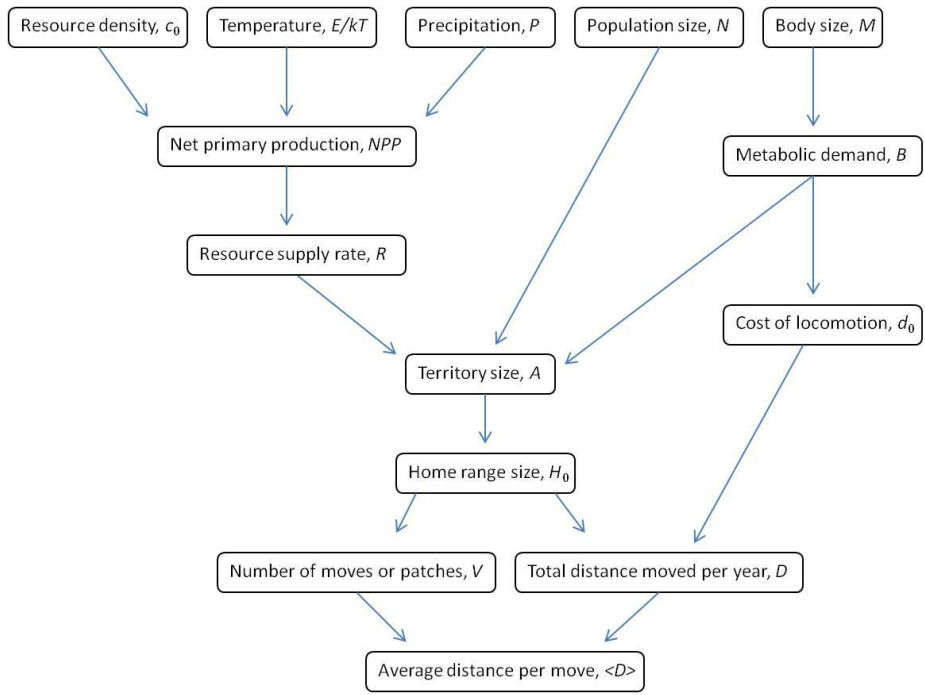

Figure 1. Flow chart of environmental, population, and individual level processes influencing hunter-gatherer space use and residential mobility examined in this paper.

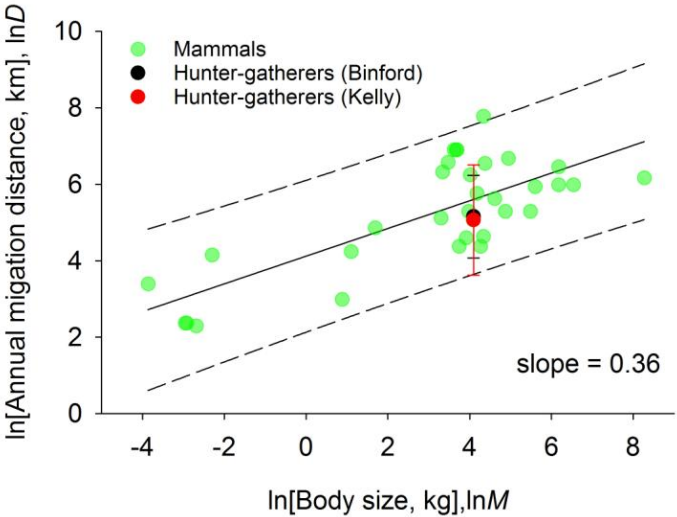

Figure 2. Annual migration distance as a function of body mass for terrestrial mammals and hunter-gatherers on logarithmic axes. Hunter-gatherer estimates are mean +/- 1 s.d. (Binford: 158.43 +/- 2.94; Kelly: 174.08 +/- 4.22 km/yr).

ln[Net primary production, g/m²/yr], ln*NPP*

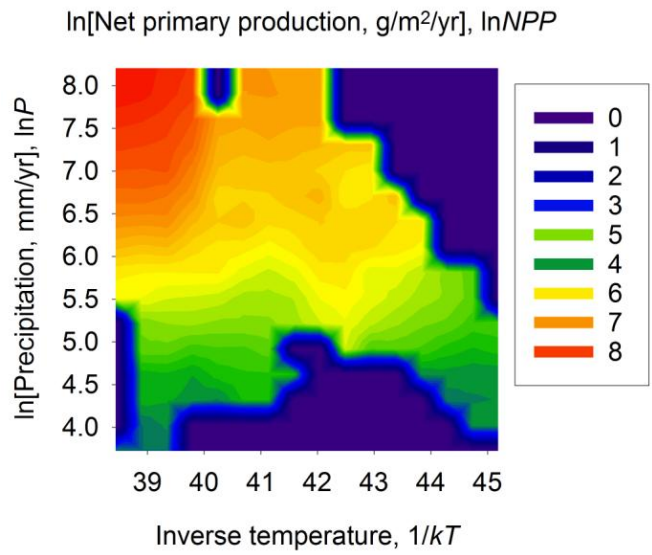

Figure 3. Net primary production as a function of inverse temperature and precipitation. High intensity reds indicate highest *NPP* and low intensity greens/blues indicate low *NPP.*

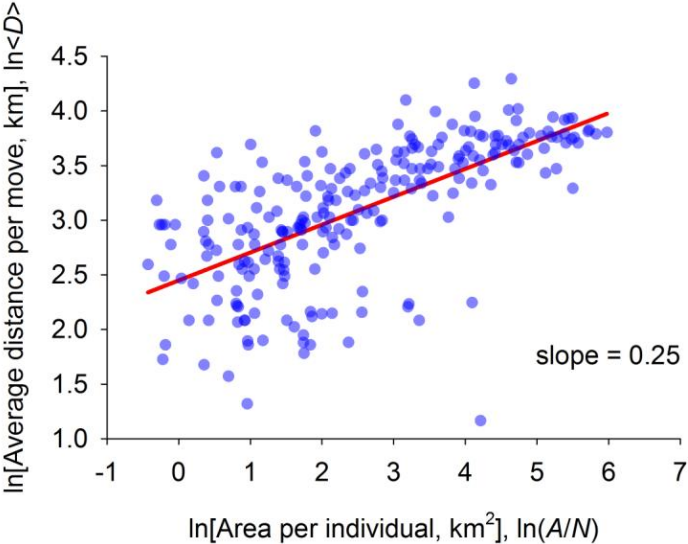

Figure 4. Average distance per move as a function of area per individual on logarithmic axes (OLS regression: $r^2 = 0.43$, $p<0.0001$, $d.f.=238$).

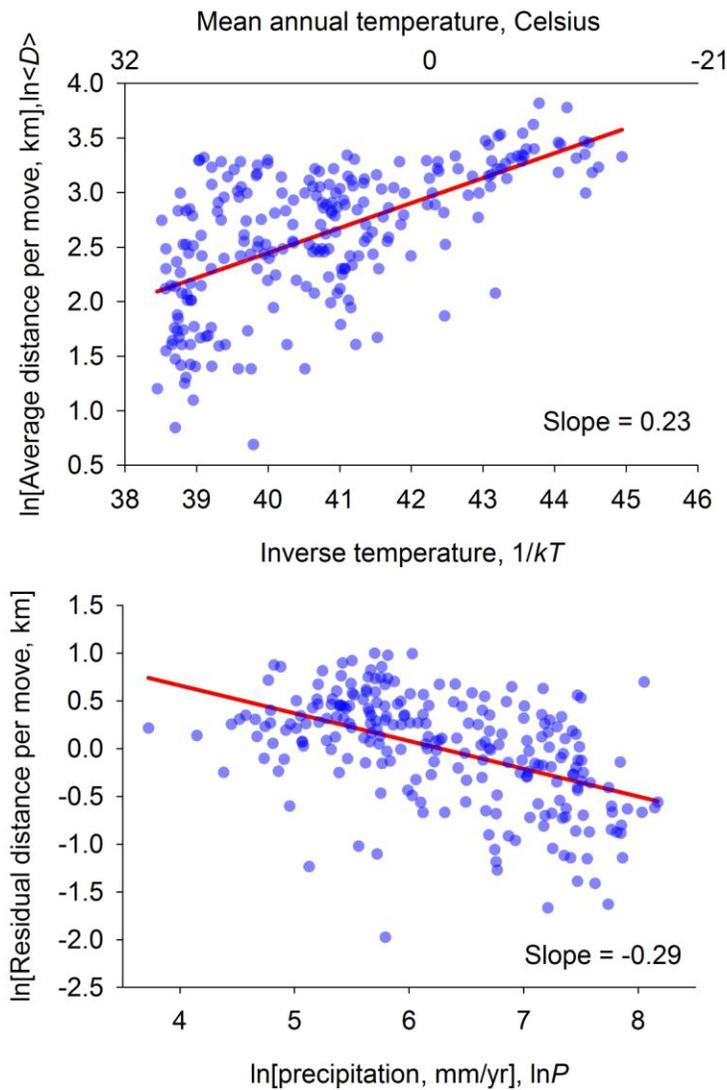

Figure 5. Average distance per move as a function of temperature and precipitation, on semi-log axes. Top panel: distance and temperature, with slope OLS regression: $r^2$=0.36, $p$<0.0001, $d.f.$=238). Bottom panel: residual distance and precipitation OLS regression: $r^2$=0.24, $p$<0.0001, $d.f.$=238)

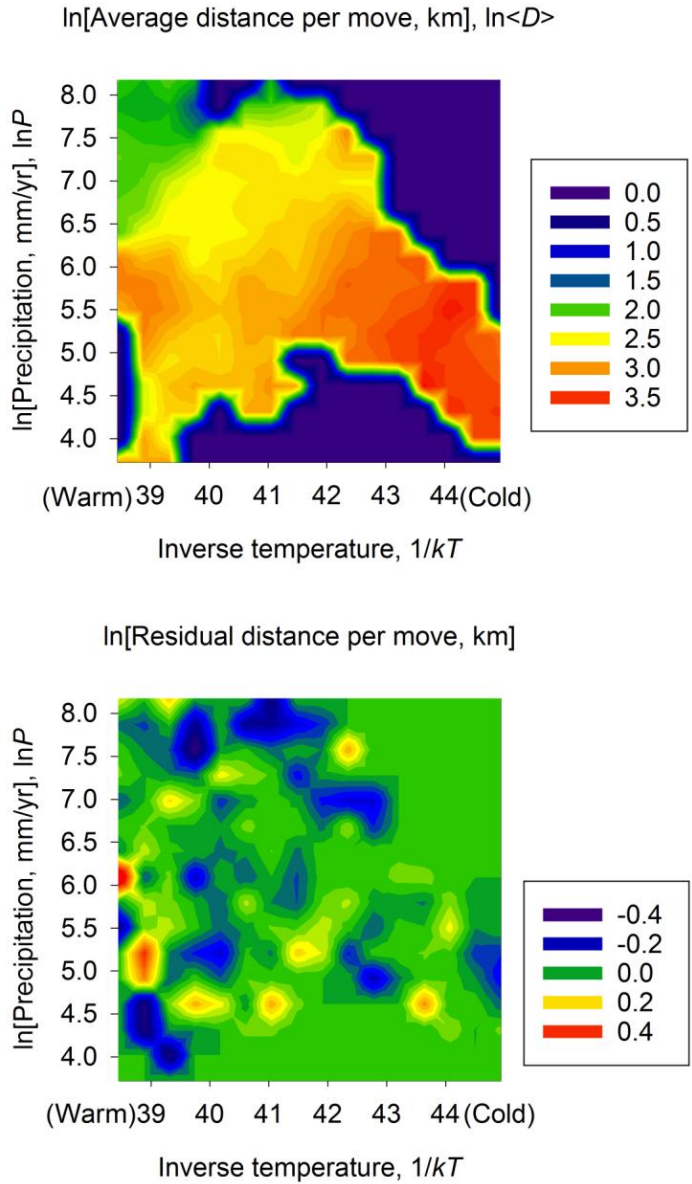

Figure 6. Contour plots of the average distance per move (top) and the residuals (bottom) as a function of annual precipitation and inverse temperature.